\documentclass[letterpaper, 10 pt, conference]{ieeeconf}
\IEEEoverridecommandlockouts                              

\usepackage[colorinlistoftodos,obeyFinal]{todonotes}                           
\usepackage{fainekos-macros}
\usepackage[]{algorithm2e}
\usepackage{amsmath} 
\usepackage{amssymb}  
\usepackage{graphicx}
\usepackage{tikz}
\usetikzlibrary{arrows}
\usepackage{ifthen}
\usepackage{mathtools}
\usepackage{stmaryrd}
\usepackage{csquotes}
\usepackage{xargs}                      
\usepackage{subcaption}
\usepackage{textcomp}
\usepackage{bm}
\usepackage{cite}
\graphicspath{{./figures/}}

\newcommand{\rob}{\rho}
\newcommand{\robf}{\rho_{\formula}}

\newcommand{\srob}{\tilde{\rob}}

\usepackage{lipsum}

\newtheorem{corollary}{Corollary}[theorem]
\newtheorem{prop}{Proposition}
\usepackage{etoolbox}
\makeatletter
\patchcmd{\maketitle}{\@copyrightspace}{}{}{}
\makeatother

\usepackage{xcolor}

\usepackage{stmaryrd}
\usepackage{hyperref}
\begin{document}



\title{Co-design of Control and Planning for Multi-rotor UAVs with Signal Temporal Logic Specifications}

%
\author{Yash Vardhan Pant$^{*}$, He Yin$^{*}$, Murat Arcak, Sanjit A. Seshia
\thanks{$^{2}${Department of Electrical Engineering and Computer Sciences, University of California, Berkeley, USA}
{\tt\footnotesize yashpant@berkeley.edu, he\_yin@berkeley.edu, arcak@berkeley.edu, sseshia@berkeley.edu}.}%
\thanks{$^*$The authors contributed equally.}
}

\maketitle

\begin{abstract}

Urban Air Mobility (UAM), or the scenario where multiple manned and Unmanned Aerial Vehicles (UAVs) carry out various tasks over urban airspaces, is a transportation concept of the future that is gaining prominence. UAM missions with complex spatial, temporal and reactive requirements can be succinctly represented using Signal Temporal Logic (STL), a behavioral specification language. However, planning and control of systems with STL specifications is computationally intensive, usually resulting in planning approaches that do not guarantee dynamical feasibility, or control approaches that cannot handle complex STL specifications. Here, we present an approach to co-design the planner and control such that a given STL specification (possibly over multiple UAVs) is satisfied with trajectories that are dynamically feasible and our controller can track them with a bounded tracking-error that the planner accounts for. 
The tracking controller is formulated for the non-linear dynamics of the individual UAVs, and the tracking error bound is computed for this controller when the trajectories satisfy some kinematic constraints.
We also augment an existing multi-UAV STL-based trajectory generator in order to generate trajectories that satisfy such constraints. We show that this co-design allows for trajectories that satisfy a given STL specification, and are also dynamically feasible in the sense that they can be tracked with bounded error. The applicability of this approach is demonstrated through simulations of multi-UAV missions. 

\end{abstract}


\section{Introduction}
\label{sec:intro}

For Urban Air Mobility to become a reality in the near future, the underlying planning and control approaches for the Unmanned Aerial Vehicles (UAVs) carrying out various tasks should be robust and have strong guarantees on safety. For this, the UAVs must respect \emph{spatial} requirements, e.g. no-fly zones around geo-fenced areas and maintaining safe distances from each other, \emph{temporal} requirements like visiting particular regions only during pre-defined time intervals, and \emph{reactive} requirements, e.g. executing a contingency plan in case of a system failure. It is hard to capture these requirements with the objectives used in traditional control and planning approaches, however recent work on planning and control with Signal Temporal Logic (STL) \cite{pant2018fly, pant2017smooth, 8404080} has shown promise in representing these missions with STL, and also on developing algorithms for UAVs and fleets of UAVs to satisfy these requirements. However in order to be computationally tractable, these approaches either work with simplifying abstractions of the system dynamics \cite{pant2018fly} or the workspace \cite{SahaRSJ14}, or cannot use the full expressiveness of STL \cite{8404080}, e.g. use of the logical \textit{or} operator to represent choices between multiple tasks in a mission.

\begin{figure}[t]
\centering
\includegraphics[width=0.45\textwidth, trim={0cm 0cm 0cm 0cm},clip]{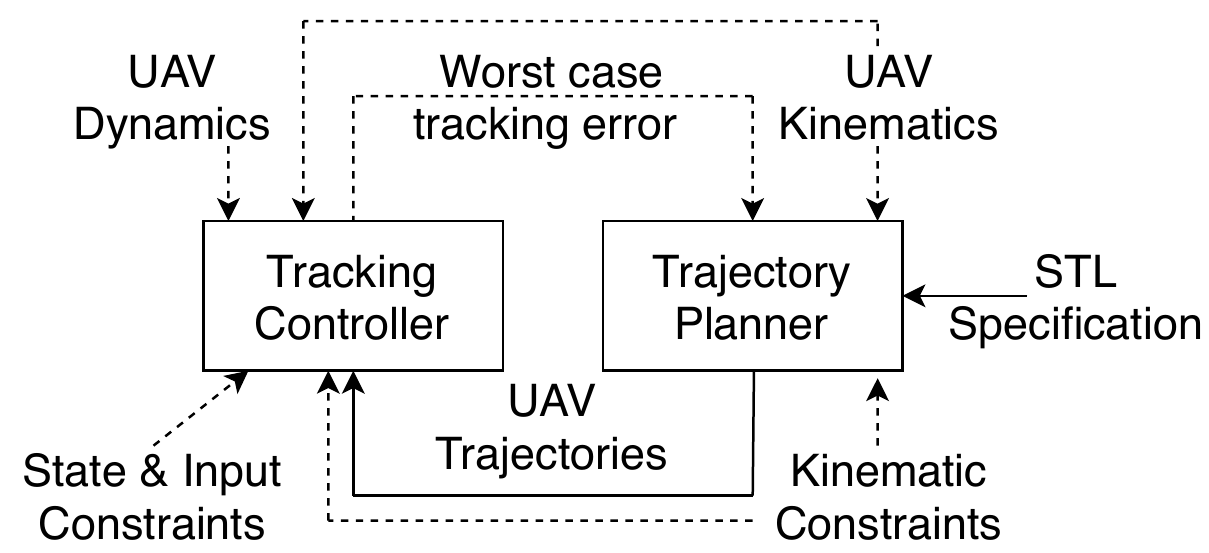}
\vspace{-5pt}
\caption{\small{Overview of the co-design framework. The dashed lines represents information used in the offline design phase, while solid line represent information used during online execution.}}
\label{fig:codesign}
\vspace{-20pt}
\end{figure}

In this paper, we present an approach that overcomes these limitations by co-designing the planning and control for multi-rotor UAVs with missions represented as STL specifications (see figure \ref{fig:codesign}). To do so, we augment a state-of-the-art trajectory planning approach \cite{pant2018fly} to generate trajectories such that a controller that considers the non-linear UAV dynamics can track them. We also use sum-of-squares programming \cite{Parrilo:00} to synthesize a controller that can track these trajectories with bounded error. We explicitly compute this bound and use it in the trajectory planner to ensure that the planned trajectories \emph{robustly} satisfy the STL specification, and that the system, when tracking with our controller, does not violate the specification. 

\vspace{-5pt}
\subsection{Contributions of this work}


The main contributions of our co-design approach are that it ensures:

\begin{enumerate}
\item The trajectories (generated via a centralized optimization) satisfy the STL specification in continuous time.
\item They also respect pre-defined kinematic constraints in the form of bounds on velocity, acceleration, and jerk.
\item The non-linear dynamics and state/input constraints of the UAVs are taken into account by the tracking controller, and worst-case tracking error bounds are computed.
\item The trajectory planner satisfies the STL specification with a robustness margin, large enough to accommodate the tracking error bound.
\end{enumerate}

Additionally, our approach can harness the full grammar of Signal Temporal Logic (see section \ref{sec:STL_intro}). Through two simulation case studies, we show how this framework can be applied to realistic scenarios and how the guarantees above translate to actual control performance and robust satisfaction of the STL specification.  


\subsection{Related work}
\label{sec:related}
\textbf{Planning and Control with STL specifications.}
The problem of controlling systems with STL specifications has been studied extensively. \cite{Raman14_MPCSTL, Sadraddini15Allerton} build upon \cite{bemporad1999control} to develop a Mixed Integer Programming-based Model Predictive Control approach (MPC). These methods are developed with Linear Time Invariant (LTI) systems in mind, and are computationally  complex to be applied for the control of multiple UAVs with complex STL specifications. \cite{pant2017smooth} presents a non-convex optimization-based MPC approach that is applicable to some non-linear systems, but is also computationally heavy to be used for online control. A barrier function-based approach is presented in \cite{8404080}, but is applicable only to a fragment of STL. Given the computational limitations in developing controllers for STL specifications, the work in \cite{pant2018fly} instead aims to generate trajectories for multi-rotor UAVs such that they satisfy a given STL specification. It however offers only kinematic feasibility of the planned trajectories, and not dynamic feasibility. 
Our co-design framework builds on this approach, and we show we can also ensure dynamic feasibility through additional constraints, and through our design of a tracking controller. There are also planning approaches that, unlike our approach, discretize the workspace and abstract away the system dynamics  (e.g. \cite{SahaRSJ14}). 

\textbf{Co-designing the planner and controller.}
A hierarchical control framework has been explored by the path planning community \cite{kousik2018bridging, herbert2017fastrack, Singh2018RobustTW}. In this framework, a low-fidelity model is used for planning online, and a high-fidelity model is used for designing a tracking controller offline, and the controller is then implemented online. A robust forward reachable set used for planning is computed in \cite{kousik2018bridging} by considering the difference between the planning and tracking models. In \cite{herbert2017fastrack}, a Hamilton-Jacobi equation based method is presented to synthesize a tracking error bound and an optimal tracking controller. An extension in \cite{Singh2018RobustTW} shows that  for a class of robotics systems, the problem can also be solved using sum-of-squares (SOS) programming. Related works other than those from the path planing literature include the continuous-state abstraction \cite{GIRARD_CA, SMITH_automatica}, which provides a tracking controller and ensure the boundedness of the error between the original high dimensional linear system and its low dimensional abstraction. This idea is extended to a general class of nonlinear systems in \cite{smithCDC, PJ_IFAC, YinAcc}, and is applied to UAVs in this paper.

\subsection{Outline of the paper}
In Section \ref{sec:problemandprelims} we formalize the problem that we aim to solve and provide a brief introduction to Signal Temporal Logic and its associated robustness metric.
Section \ref{sec:overview} gives an overview of the co-design approach, introduces the system dynamics used for planning and control, and the associated hierarchical control structure. Section \ref{sec:planning} presents the trajectory planner, which builds upon the work in \cite{pant2018fly}, and discusses how the controller performance and kinematic constraints are taken into account. Section \ref{sec:controller} presents the synthesis for the tracking controller that uses the full non-linear UAV dynamics, and its associated tracking error. Two case studies for multi-UAV settings with STL specifications are presented in Section \ref{sec:simulations}. Finally, we conclude with a discussion of the limitations of this approach and future work in Section \ref{sec:conclusion}.

\subsection{Notations}
For $\xi \in \mathbb{R}^n$, $\mathbb{R}[\xi]$ represents the set of polynomials in $\xi$ with real coefficients, and $\mathbb{R}^m[\xi]$ and $\mathbb{R}^{m \times p}[\xi]$ to denote all vector and matrix valued polynomial functions. The subset $\Sigma[\xi] := \left\{\pi = \sum_{i=1}^M \pi_i^2 : M\ge 1, \pi_i \in \mathbb{R}[\xi]\right\}$ of $\mathbb{R}[\xi]$ is the set of sum-of-squares (SOS) polynomials. The state of UAV $i$ is denoted by $x_i \in \mathbb{R}^n$. A trajectory of states over a time interval $[0,T]$ is denoted by $\mathbf{x}_i : [0,T] \rightarrow \Re^n$. The value of the state at time $t$ is denoted by $\mathbf{x}_i(t)$.
\section{Problem Statement and preliminaries}
\label{sec:problemandprelims}
\subsection{Problem Statement} 
\label{sec:problem}
Formally, in order to achieve the objectives laid out earlier in the paper, we aim to solve the following problem:

\begin{prob}
\label{prob:overall}
Given UAV dynamics $\dot{\mathbf{x}}_i(t) = f(\mathbf{x}_i(t)) + g(\mathbf{x}_i(t))\mathbf{u}_i(t)$ and output $\mathbf{y}_i(t) = C\mathbf{x}_i(t)$ for UAV $i\in\{1,\dotsc,D\}$, control the UAVs such that resulting output trajectories ($\mathbf{y}_i$) are such that:
\begin{enumerate}
\item $(\mathbf{y}_1,\dotsc,\mathbf{y}_D) \models \formula(\mathbf{y}_1,\dotsc,\mathbf{y}_D)$, i.e. satisfy the STL specification $\formula$ and,
\item respect constraints $\mathbf{x}_i(t) \in X \subset \mathbb{R}^n,\, \mathbf{u}_i(t) \in U \subset \mathbb{R}^m, \ \forall i=1,\dotsc,D, \, \forall t \in [0,\text{hrz}(\formula)]$.
\end{enumerate}
\end{prob}

Here, $\mathbf{x}_i(t)$ and $\mathbf{u}_i(t)$ are the state and input for UAV $i$ at time $t$, $X$ and $U$ are (hyper-rectangular) state and input constraints, and $\text{hrz}(\formula)<\infty$ is the time horizon of the specification $\formula$, i.e. the minimum time duration needed to evaluate $\formula$. For simplicity we assume that all UAVs have identical dynamics, however this not necessary for our approach to work. 

In order to avoid the pitfalls of existing approaches, we solve this problem by solving the two following sub-problems instead, which constitute a hierarchical planner-controller framework:

\begin{prob}[Trajectory planning]
\label{prob:planner}
Develop a planner that uses the UAV kinematics of the form $\dot{\hat{\mathbf{x}}}_i(t) = A\mathbf{\hat{x}}_i(t) + B\mathbf{\hat{u}}_i(t)$ (details in Section \ref{sec:kinematics}) to generate trajectories $\mathbf{\hat{x}_i}$ such that:
\begin{enumerate}
\item $(\mathbf{\hat{x}}_1,\dotsc,\mathbf{\hat{x}}_D) \models \formula(\mathbf{\hat{x}}_1,\dotsc,\mathbf{\hat{x}}_D)$ and,
\item $\mathbf{\hat{x}}_i(t) \in \hat{X},\, \mathbf{\hat{u}}_i(t) \in \hat{U}\, \forall i = 1,\dotsc,D, \, \forall t \in [0,\text{hrz}(\formula)]$,
\item any trajectories $\mathbf{x}_i$ where $||C(\mathbf{x}_i(t)-G\mathbf{\hat{x}}_i(t))||_\infty \leq \delta, \ \forall t \in [0,\text{hrz}(\formula)]$ also satisfy $\formula$.
\end{enumerate}

\end{prob}

Here, $G$ is a matrix of appropriate size that relates the states in the kinematic model to those in the full dynamics, and $\delta \in \mathbb{R}_{>0}$ is the worst-case tracking error bound. See Section \ref{sec:controller} for details. The controller design is done individually for each UAV, so we drop the subscript $i$ for ease of notation:

\begin{prob}[Tracking Controller]
\label{prob:controller}
For the full UAV dynamics (problem \ref{prob:overall}), develop a controller $\mathbf{u}(t) = k(\mathbf{e}(t))$ s.t. the error state $\mathbf{e}(t) = \mathbf{x}(t)-G\mathbf{\hat{x}}(t)$ for tracking a trajectory $\mathbf{\hat{x}}$ is s.t. $||\mathbf{e}(t)||_\infty \leq \delta$, and $\mathbf{x}(t)\in X,\, \mathbf{u}(t) \in U\, \ \forall t\in[0,\text{hrz}(\formula)]$.
\end{prob}

\subsection{Review of Signal Temporal Logic (STL)}
\label{sec:STL_intro}

Signal Temporal Logic (STL) \cite{MalerN2004STL} is a behavioral specification language that can be used to encode requirements on signals. 
The grammar of STL~\cite{Raman14_MPCSTL} allows for capturing a rich set of 
behavioral requirements using temporal operators, such as the \textit{Until} ($\until$) operator and derived \textit{Always} ($\always$) and \textit{Eventually} ($\eventually$), 
 as well as logical operators \textit{And} ($\land$), \textit{Or} ($\lor$), and \textit{negation} ($\neg$). 
With these operators, an STL specification $\varphi$ is defined over a signal, e.g. over the trajectories of the UAVs, and evaluates to either \textit{True} or \textit{False}. 

\begin{figure}[tb]
\centering
\includegraphics[width=0.49\textwidth, trim={2cm 1cm 2cm 1cm},clip]{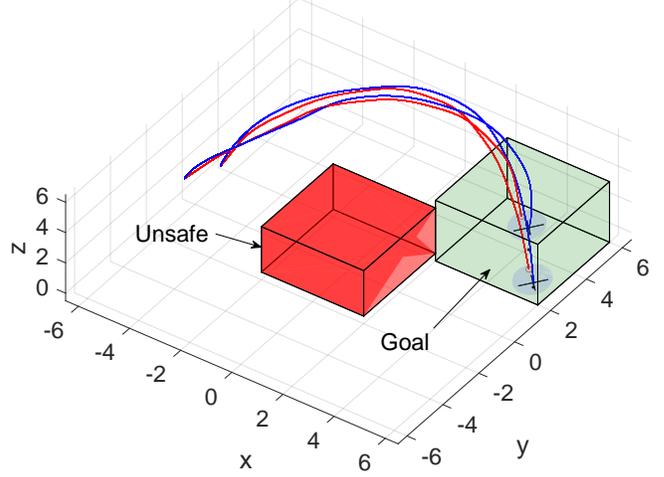}
\vspace{-10pt}
\caption{\small{The workspace, planned trajectories (red) and the tracked trajectories (blue) for the two UAVs carrying out the timed reach-avoid mission of example \ref{ex:reach_avoid_exmp}. Simulation videos are at \protect\url{http://bit.ly/TimedRA}. Color in digital version.}}
\label{fig:RA}
\vspace{-5pt}
\end{figure}

\begin{exmp}
\label{ex:reach_avoid_exmp}
\textit{(A two-UAV timed reach-avoid problem)} 
Two quad-rotor UAVs are tasked with a mission with spatial and temporal requirements in the workspace shown in Fig. \ref{fig:RA}:

\begin{enumerate}

\item The two UAVs have to reach a $\text{Goal}$ set (shown in green), 
or a region of interest, within a time of $8$ seconds after starting. 
UAV $j$ (where $j\in \{1,2\}$), with position (in 3D) denoted by $p_j$, has to satisfy: $\varphi_{\text{reach}, j} = 
\eventually_{[0,8]} (p_j \in \text{Goal})$.
The \textit{Eventually} operator over the time interval $[0,8]$ requires UAV $j$ to be inside the set $\text{Goal}$ at some point within $8$ seconds. 

\item The two UAVs also have to avoid an $\text{Unsafe}$ set (in red), or no-fly zone. For each UAV $j$, this is encoded with \textit{Always} and \textit{Negation} operators:

$\varphi_{\text{avoid},j} = \always_{[0,8]} \neg (p_j \in 
\text{Unsafe})$

\item Finally, the two UAVs should also be separated by at least $0.2$ meters along every axis of motion:

$\varphi_{\text{separation}} = \always_{[0,8]} ||p_1 - p_2||_{\infty} 
\geq 0.2$

\end{enumerate}

These requirements are combined into a two UAV timed reach-avoid specification: 
\vspace{-7pt}
\begin{equation}
\label{eq:timed_RA}
\varphi_{\text{reach-avoid}} = \land_{j=1}^2 ( \varphi_{\text{reach},j} \land 
\varphi_{\text{avoid},j}) \land \varphi_\text{separation}
\end{equation}
\end{exmp}

In order to satisfy $\varphi$, a planning method generates trajectories $\mathbf{\hat{x}}_1, \mathbf{\hat{x}}_2$ (where position $p$ is obtained from the full kinematic state $\hat{x}$ as $p=Cx$) of a duration that is at least $\text{hrz}(\varphi)= 8$s, where $\text{hrz}(\varphi)$ is the time \textit{horizon} of $\varphi$. 
If the trajectories satisfy the specification, i.e. $(\mathbf{p}_1,\, \mathbf{p}_2) \models \varphi$, then the specification $\varphi$ evaluates to \textit{True}, otherwise it is \textit{False}. 
In general, an upper bound for the time horizon can be computed as shown in \cite{Raman14_MPCSTL}. 
Here, we only consider specifications for which the horizon is bounded. More details on STL are in \cite{MalerN2004STL} or \cite{Raman14_MPCSTL}. 

\subsection{Robustness of STL specifications}
\label{sec:stl_rob_short}

For every STL specification $\formula$, we can construct a \textit{robustness function} \cite{FainekosP09tcs} by following the grammar of STL. It outputs a \textit{robustness} value $\rho_\formula(\mathbf{x})$ for this formula, with respect to the signal $\mathbf{x}$ that it is defined over, and has the important following property:

\begin{theorem} \cite{FainekosP09tcs}
	\label{thm:rob objective new}
	For any $\sstraj$ and STL formula $\formula$, 	
	if $\robf(\sstraj) <0$ then $\sstraj$ violates $\formula$, and if $\robf(\sstraj) > 0$ then $\sstraj$ satisfies $\formula$. 
	The case $\robf(\sstraj) =0$ is inconclusive.
\end{theorem} 
Thus, the degree of satisfaction or violation of a specification is indicated by the robustness value.
For simplicity, the  distances are defined in the inf-norm sense. 
This, combined with Theorem \ref{thm:rob objective new} gives us the following result:

\begin{corollary}
\label{cor:rob_tube}
Given a trajectory $\mathbf{\hat{x}}$ such that $\mathbf{\hat{x}} \models \formula$ with robustness value $\rho_\formula(\mathbf{\hat{x}})>0$, then any trajectory $\mathbf{x}$ that is within $\rho_\formula(\mathbf{\hat{x}})$ of $\mathbf{\hat{x}}$ at each time, i.e. $||\mathbf{\hat{x}}(t)-\mathbf{x}(t)||_\infty < \rho_\formula(\mathbf{\hat{x}}) \, \forall t \in [0,\text{hrz}(\formula)]$, is such that $\mathbf{x} \models \formula$ (satisfies $\formula$).
\end{corollary}

\noindent \textbf{Significance:} We use this property in the co-design, where the planner generates a trajectory $\mathbf{\hat{x}}$ and the tracking controller tracks it with a trajectory $\mathbf{x}$ such that $||C(\mathbf{x}(t)-G\mathbf{\hat{x}}(t))||_\infty \leq \rho_\formula(\hat{\mathbf{x}}) \,\forall t \in [0,\text{hrz}(\formula)]$. This requires the the inf-norm of the worst-case tracking error is always less than $\rho_\formula(\mathbf{\hat{x}})$.  Note, $\formula$ is usually over a subset of states given by $Cx$, e.g. the specification \ref{eq:timed_RA} is over the position components of the UAV state. In general, our approach allows for specifications over the translational positions, velocities and accelerations of the UAV (Section \ref{sec:planning}).

\section{Overview of the Co-design approach}
\label{sec:overview}

\begin{figure}[tb]
\centering
\includegraphics[width=0.49\textwidth, trim={0cm 9.75cm 0cm 1.2cm},clip]{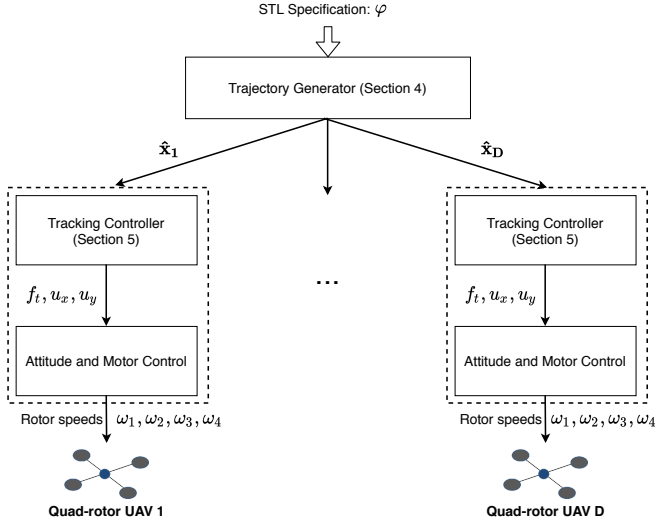}
\vspace{-10pt}
\caption{\small{The planning and control hierarchy for multi UAV missions.}}
\label{fig:hierarchy}
\vspace{-5pt}
\end{figure}

In order to solve problems \ref{prob:planner} and \ref{prob:controller}, we adopt a hierarchical planning and control stack for the UAVs of the kind that is commonly used, see figure \ref{fig:hierarchy}. Our framework co-designs (see figure \ref{fig:codesign}) the two main components here, namely:

\begin{itemize}
\item \textbf{Trajectory planner} that takes as input the STL specification defining the (possibly multi-UAV) mission. Using the UAV kinematics (Section \ref{sec:kinematics}, the trajectory generator solves an optimization (in a centralized manner) that generates trajectories for all UAVs involved. The trajectories satisfy the STL specification in continuous time, ensure that kinematic constraints are respected, and take into account the worst-case tracking error. As long as the tracking controller does not deviate from a given bound, obtained from this optimization's objective, the tracking of these trajectories will be such that the STL specification is satisfied. Section \ref{sec:planning} presents the details. 

\item \textbf{Tracking controller} that follows the trajectories generated by the planner by taking into account both the non-linear dynamics of multi-rotor UAVs (Section \ref{sec:NLdynamics}) and the kinematics that the planner uses to generate trajectories. This controller is synthesized via a sum-of-squares optimization through which we obtain (state-dependent) tracking error bounds and worst-case tracking error bounds, which are then used in the trajectory planning optimization. Details on the controller design are in Section \ref{sec:controller}.

\end{itemize}

Commands from the tracking controller are realized by lower level attitude controllers (Section \ref{sec:lowerlevel}). The details of attitude controllers are beyond the scope of this paper, but standard approaches can be found in \cite{sabatino2015quadrotor}. The resulting planning and control hierarchy is shown in figure \ref{fig:hierarchy}. Note that the trajectory planning is done in a centralized manner, but the synthesized tracking controllers run independently on each UAV to track these generated trajectories. The rest of this section introduces the UAV kinematics and dynamics used by the trajectory planner and tracking controller, respectively.



\subsection{Multi-rotor UAV kinematics for trajectory planning}
\label{sec:kinematics}
To generate UAV trajectories that satisfy a STL specification $\formula$, we use the method in \cite{pant2018fly} that generates trajectories by connecting segments of jerk-minimizing splines \cite{MuellerTRO15}. These are based on the following kinematic model.

{\small
\begin{equation}
\label{eq:9state_system}
    \begin{split}
        \dot{\hat{\mathbf{x}}}(t) = \begin{bmatrix}  \dot{\hat{\mathbf{p}}}_x(t) \\ \dot{\hat{\mathbf{p}}}_y(t) \\ \dot{\hat{\mathbf{p}}}_z(t) \\ \dot{\hat{\mathbf{v}}}_x(t) \\ \dot{\hat{\mathbf{v}}}_y(t) \\ \dot{\hat{\mathbf{v}}}_z(t) \\ \dot{\hat{\mathbf{a}}}_x(t) \\\dot{\hat{\mathbf{a}}}_y(t) \\\dot{\hat{\mathbf{a}}}_z(t)\\ \end{bmatrix} =
        \begin{bmatrix} \hat{\mathbf{v}}_x(t) \\ \hat{\mathbf{v}}_y(t) \\ \hat{\mathbf{v}}_z(t)\\ \hat{\mathbf{a}}_x(t) \\ \hat{\mathbf{a}}_y(t) \\ \hat{\mathbf{a}}_z(t) \\ 0 \\ 0 \\ 0 \\ \end{bmatrix}
        + \begin{bmatrix} 0 \\ 0 \\ 0 \\ 0 \\ 0 \\ 0 \\ \hat{\mathbf{j}}_x(t) \\ \hat{\mathbf{j}}_y(t)\\ \hat{\mathbf{j}}_z(t) \end{bmatrix}
    \end{split}
\end{equation}}

Here $\hat{p}_x$, $\hat{v}_x$, $\hat{a}_x$ represents the position, velocity, and acceleration of the UAV in the global $x$ coordinate (and similarly for $y$ and $z$ coordinates), and jerks $\hat{j}_x$, $\hat{j}_y$, and $\hat{j}_z$ are the inputs $\hat{u}$ to this system.

\subsection{Quad-rotor UAV dynamics}
\label{sec:NLdynamics}

For the UAVs, we consider the following non-linear dynamics, adapted from \cite[Chapter 2]{sabatino2015quadrotor},
{\small
\begin{equation}
\label{eq:8state_system}
    \begin{split}
        \dot{\mathbf{x}}(t) = \begin{bmatrix} \dot{\mathbf{p}}_x(t) \\ \dot{\mathbf{p}}_y(t) \\ \dot{\mathbf{p}}_z(t)\\ \dot{\mathbf{v}}_x(t) \\ \dot{\mathbf{v}}_y(t) \\ \dot{\mathbf{v}}_z(t) \\ \dot{\text{\boldmath$\phi$} }(t) \\ \dot{\text{\boldmath$\theta$} }(t) \end{bmatrix} =
        \begin{bmatrix} \mathbf{v}_x(t) \\ \mathbf{v}_y(t) \\ \mathbf{v}_z(t) \\ -\frac{\mathbf{f}_t(t)}{m}c_{\text{\boldmath$\phi$}(t)} s_{\text{\boldmath$\theta$}(t)} \\ \frac{\mathbf{f}_t(t)}{m}s_{\text{\boldmath$\phi$}(t)} \\ g-\frac{\mathbf{f}_t(t)}{m}c_{\text{\boldmath$\phi$}(t)} c_{\text{\boldmath$\theta$}(t)} \\ \frac{\mathbf{u}_x(t)}{I_x} \\ \frac{\mathbf{u}_y(t)}{I_y},  \end{bmatrix}
    \end{split}
\end{equation}}
\hspace{-8pt} where $c_\phi := \cos(\phi)$, $s_\phi := \sin(\phi)$ (and similarly for $\theta$).

The 8-state system has 6 translational states $p_x$, $p_y$ and $p_z$ (the positions of the quad-rotor in a global co-ordinate frame) and $v_x$, $v_y$ and $v_z$ (velocities), as well as two states for orientation $\phi$ and $\theta$ which are the roll and pitch angles respectively. To obtain this model, we assume that yaw $\psi$ and its derivatives are zero throughout. The control inputs to this model are $f_t$ (thrust), and $u_x$ and $u_y$ which are related to the torques generated by the rotors. We define a variable $u$ to gather all the control inputs: $u = [f_t, u_x, u_y]^\top$. Without loss of generality, we assume the input constraint set is of the form: $U = \{u \in \Re^{3}: \underline{u} \leq u \leq \overline{u} \}$, where $\underline{u}, \overline{u} \in \Re^3$, and $\leq$ is applied elementwise. Here, $m$ is the UAV's mass, and $I_x$ and $I_y$ are the moments of inertia with respect to $x$ and $y$ axes in the body-fixed frame. Values for them are: $m = 0.5$, and $I_x = I_y = 0.2$. \footnote{For simplicity, we assume all UAVs have identical dynamics, however our approach would apply otherwise as well.}

\subsection{Error dynamics for the tracking controller}

To help keep track of the ``distance'' between the planning and tracking trajectories, we define the error state as:

{\small
\begin{align}
    e = x- G \hat{x} = \begin{bmatrix} p_x - \hat{p}_x \\ p_y - \hat{p}_y \\ p_z - \hat{p}_z \\ v_{x} - \hat{v}_x \\ v_{y} - \hat{v}_y \\ v_{z} - \hat{v}_z \\ \phi \\ \theta\end{bmatrix},
\end{align}}
\hspace{-3pt}which include the positional and velocity differences, and the roll and pitch angles of the UAV. Here $G = \text{diag}(I_6, 0_{2 \times 3})$. Later in Section~\ref{sec:controller}, we will design a tracking controller to penalize the error state $e$, and ensure its boundedness.

The corresponding error dynamics are:
\begin{align}
    \dot{\mathbf{e}}(t) &= f_e(\mathbf{e}(t), \hat{\mathbf{x}}(t)) + g_e(\mathbf{e}(t))\mathbf{u}(t), \label{eq:err_dyn}  
\end{align}
    where
   {\small
\begin{align}
    f_e(e,\hat{x}) &= \begin{bmatrix} e_4 \\ e_5 \\ e_6 \\  - \hat{a}_x \\  - \hat{a}_y \\ g - \hat{a}_z \\ 0 \\ 0 \end{bmatrix},  \
    g_e(e) = \begin{bmatrix} 0 &0 &0\\0 &0 &0\\0 &0 &0\\ \frac{-c_{e_7} s_{e_8}}{m}  &0 &0 \\ \frac{s_{e_7}}{m} &0 &0 \\ \frac{-c_{e_7} c_{e_8}}{m} &0 &0 \\ 0 & \frac{1}{I_x} &0 \\ 0 &0 &\frac{1}{I_y}\end{bmatrix}.\nonumber
\end{align}}
\hspace{-3pt}In the error dynamics, $\hat{a}_x, \hat{a}_y$ and $\hat{a}_z$ are treated as bounded disturbances, whose bounds are enforced by the STL based planning algorithm described in Section~\ref{sec:planning}. 


\subsection{Lower level attitude and motor controllers}
\label{sec:lowerlevel}
The attitude controller takes as measurements the roll and pitch rates $\dot{\phi}, \, \dot{\theta}$, and $u_x, \, u_y$ as references. It is tasked with generating torques $T_x$, $T_y$ and $T_z$ to realize $u_x$ and $u_y$.

{\small
\begin{equation}
\label{eq:attitude_controller_dynamics}
    \begin{split}
        \dot{\mathbf{u}}_x(t) &= \mathbf{T}_x(t) \\
        \dot{\mathbf{u}}_y(t) &= \mathbf{T}_y(t) \\
        \mathbf{T}_z(t) &= (I_y-I_x)\dot{\text{\boldmath$\phi$}}(t)\dot{\text{\boldmath$\theta$}}(t) \\
    \end{split}
\end{equation}}

Finally, from the computed values of $f_t$, $T_x$, $T_x$ and $T_z$, the desired rotor (angular) speeds $\omega_i,\, i\in\{1,\dotsc,4\}$ can be computed by solving the equality constrained system \cite{sabatino2015quadrotor}:

{\small
\begin{equation}
    \begin{split}
        f_t &= b\sum_{i=1}^4 \omega_i^2 \\
        T_x &= bl(\omega_3^2 - \omega_1^2)\\
        T_y &= bl(\omega_4^2 - \omega_2^2)\\
        T_z &= d(\omega_4^2 +  \omega_2^2 - \omega_3^2 - \omega_1^2)
    \end{split}
\end{equation}}

Here, $b$, $l$, $d$ are UAV specific constants \cite{sabatino2015quadrotor}. The solution to this system acts as the desired angular speeds for the lower-level motor controllers to track. The formulation of these lower-level controllers is beyond the scope of this work, and the interested reader can refer to \cite{sabatino2015quadrotor} for details.

\section{Planning for UAVs with STL specifications}
\label{sec:planning}

\begin{figure}[tb]
\centering
\includegraphics[width=0.25\textwidth, trim={3cm 6.25cm 2cm 5.45cm},clip]{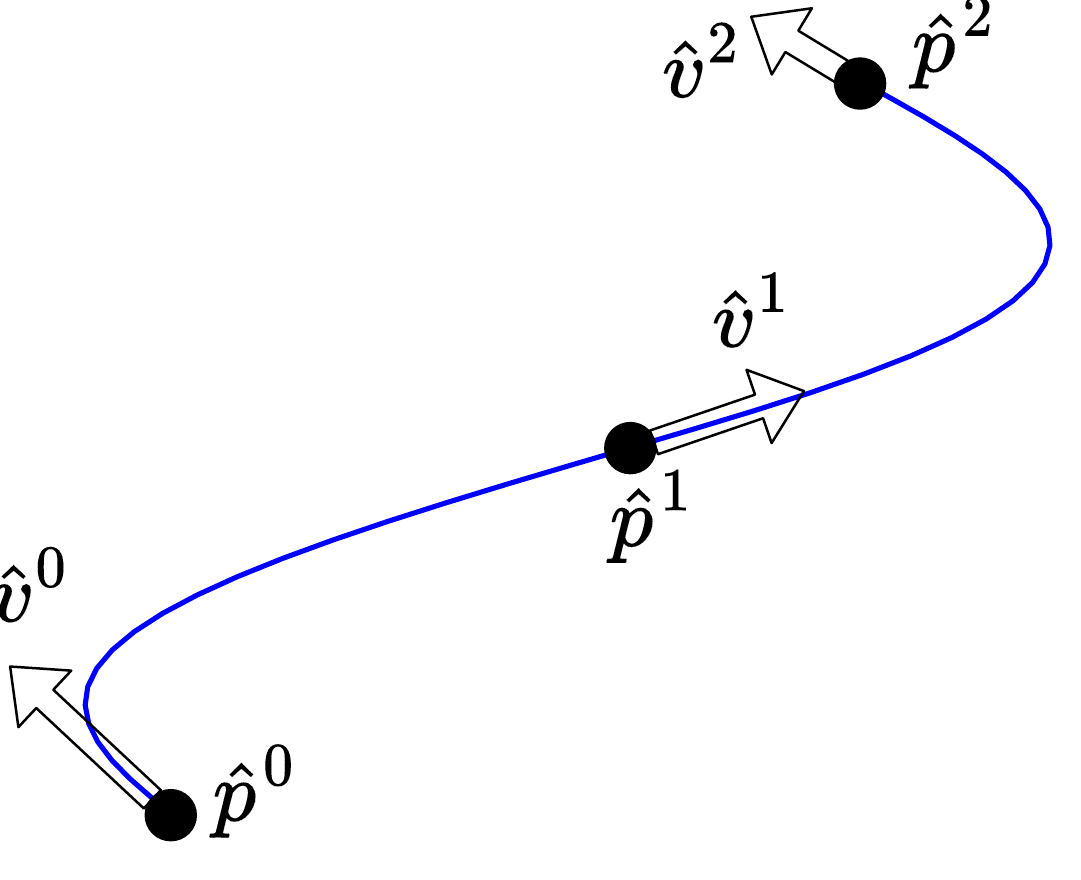}
\vspace{-10pt}
\caption{\small{Jerk minimizing splines connecting position waypoints $\hat{p}^1$, $\hat{p}^2$ and $\hat{p}^3$. The arrows show the velocities at these waypoints $\hat{v}^i,\, \forall i \in \{0,1,2\}$.  }}
\label{fig:splines}
\vspace{-20pt}
\end{figure}

For trajectory planning with a given STL specification $\formula$, defined possibly over trajectories of multiple UAVs, we use the method of \cite{pant2018fly} with some modifications. This approach maximizes the robustness $\rho_\formula$ (see Theorem \ref{thm:rob objective new}) of the given STL specification by selecting position waypoints connected by jerk minimizing splines \cite{MuellerTRO15}. The segment between one waypoint, given by position $\hat{p}^0=[\hat{p}^0_x,\hat{p}^0_y,\hat{p}^0_z]^\top$ and velocity $\hat{v}^0 = [ \hat{v}^0_x, \hat{v}^0_y, \hat{v}^0_z]^\top$, and another with desired position $\hat{p}^1 =[\hat{p}^1_x,\hat{p}^1_y,\hat{p}^1_z]^\top$, is a trajectory (see figure \ref{fig:splines}) of fixed time duration $T$ with the states of the kinematic model \eqref{eq:9state_system} ($\forall l \in \{x,y,z\}, \, \forall t \in [0,T]$) given by:
\begin{equation}
\label{eq:minjerk}
\begin{bmatrix} \mathbf{\hat{p}}_l(t) \\ \mathbf{\hat{v}}_l(t) \\ \mathbf{\hat{a}}_l(t) \end{bmatrix}
= \begin{bmatrix} \frac{\alpha}{120}t^5 + \frac{\beta}{24}t^4 + \frac{\gamma}{6}t^3 + \hat{v}_l^0t + \hat{p}_l^0 \\ \frac{\alpha}{24}t^4 + \frac{\beta}{6}t^3 + \frac{\gamma}{2}t^2 + \hat{v}_l^0 \\ \frac{\alpha}{6}t^3 + \frac{\beta}{2}t^2 + \gamma t  \end{bmatrix}
\end{equation}

Here, $\alpha$, $\beta$ and $\gamma$ are linear functions of $\hat{p}^0$, $\hat{v}^0$ and $\hat{p}^1$ (and parameter $T$) \cite{MuellerTRO15}. We assume that the start and end accelerations are zero, i.e. $\hat{a}^0 = \hat{a}^1 = 0_{3 \times 1}$ and end velocity $\hat{v}^1$ is not fixed. For brevity, we omit further details here. The interested reader can refer to \cite{pant2018fly, MuellerTRO15}.

\noindent \textbf{Trajectory planning optimization:}
We use the Fly-by-Logic method \cite{pant2018fly} to generate a sequence of $N+1$ position waypoints in 3-D space for each UAV ($i$) $\hat{p}_i^{0:N} = [\hat{p}_i^0,\dotsc,\hat{p}_i^N]$, where total flight time $NT \geq \text{hrz}(\formula)$, such that the resulting trajectories (see figure \ref{fig:splines}) of \eqref{eq:minjerk} $\mathbf{\hat{p}}_i:[0,NT] \rightarrow \mathbb{R}^9$ maximize a continuously differentiable approximation ($\srob_\formula$) \cite{pant2017smooth} of the robustness $\rho_\formula$.
\vspace{-10pt}

{\small
	\begin{subequations}
		\label{eq:HLOptim}
		\begin{align}
		&\max_{\hat{p}_1^{0:N},\dotsc,\hat{p}_D^{0:N}} \quad \srob_{\formula}([\mathbf{\hat{p}}_1,\dotsc,\mathbf{\hat{p}}_D])\\
		\text{s.t. }& \forall i=1,\dotsc,D,\, \forall j=0,\dotsc,N-1 \\		
		&\text{LB}_\text{vel}(\hat{v}^j_i) \leq \hat{p}^{j+1}_i - \hat{p}^{j}_i  \leq \text{UB}_\text{vel}(\hat{v}^j_i)  
		\label{eq:HL vel constraint}\\
		&\text{LB}_\text{acc}(\hat{v}^j_i) \leq \hat{p}^{j+1}_i - \hat{p}^{j}_i  \leq \text{UB}_\text{acc}(\hat{v}^j_i) 
		\label{eq:HL acc constraint}\\
		&\text{LB}_\text{jerk}(\hat{v}^j_i) \leq \hat{p}^{j+1}_i - \hat{p}^{j}_i  \leq \text{UB}_\text{jerk}(\hat{v}^j_i) 
		\label{eq:HL jerk constraint}
		\\
		&\srob_{\formula}([\mathbf{\hat{p}_1},\dotsc,\mathbf{\hat{p}_D}]) \geq \tilde{\epsilon} + \delta
		\label{eq:HL rob constraint}
		\end{align}
	\end{subequations}
	\vspace{-14pt}
}

This is a non-convex optimization, with linear constraints \cite{pant2018fly} and can be solved using off-the-shelf solvers like IPOPT. The UAV trajectories $\mathbf{\hat{p}_i}$ , when discretized in time, are linear functions of the position waypoints $\hat{p}_i^{0:N}$ (details in \cite{pant2018fly}). Constraints \eqref{eq:HL vel constraint}, \eqref{eq:HL acc constraint} are linear functions of the waypoints (and associated waypoint velocities) to ensure that the velocity and acceleration for each axis of motion are within predefined intervals $[v_\text{min},v_\text{max}]$ and $[a_\text{min},a_\text{max}]$ respectively. These constraints are from \cite{pant2018fly}. Additionally we develop constraints \eqref{eq:HL jerk constraint}, also linear in the optimization variables, for the jerk to be within pre-defined intervals $[j_\text{min},j_\text{max}]$ as well. The expressions for these constraints are in the Appendix. 
Finally, the constraint \eqref{eq:HL rob constraint}, where $\tilde{\epsilon}$ is the worst case approximation error \cite{pant2017smooth} of $\srob_\formula$ with respect to $\rob_\formula$, is to ensure that the STL robustness of the trajectories with respect to the specification $\formula$ is above a threshold $\delta>0$. As we will see in the next section, $\delta$ is the worst case tracking error achieved by the tracking controller. 

\begin{theorem}[STL satisfaction and kinematic feasibility]
\label{th:FbL_result}
A feasible solution to the optimization \eqref{eq:HLOptim} generates trajectories $\mathbf{\hat{p}_1},\dotsc,\mathbf{\hat{p}_D}$ such that they:
\begin{enumerate}
\item Satisfy the STL specification $\formula$ in continuous time
\item Have bounded velocity, acceleration and jerk (along every axis of motion $l \in \{x,y,z\}$) such that $\forall t \in [0, \text{hrz}(\formula)]$:  $\mathbf{\hat{v}}_l(t) \in [v_\text{min},v_\text{max}]$, $\mathbf{\hat{a}}_l(t) \in [a_\text{min},a_\text{max}]$ and $\mathbf{\hat{j}}_l(t)\in [j_\text{min},j_\text{max}]$ for each UAV.
\item As long as each UAV tracks these trajectories with tracking error $\mathbf{e}(t)$ such that $||\mathbf{e}(t)||_\infty \leq \delta \, \ \forall t \in [0,\text{hrz}(\formula)]$, the specification $\formula$ is satisfied. 
\end{enumerate}

\end{theorem}

 This trajectory planning approach hence serves as a solution to problem \ref{prob:planner}. The proofs for the first two points (except the bounded jerk constraints) can be found in \cite{pant2018fly}. The third point is a direct consequence of Corollary \ref{cor:rob_tube}.

\section{Controller and Error Bound Synthesis}
\label{sec:controller}
In \eqref{eq:err_dyn}, we have seen that the planner states enter the error dynamics. When designing a tracking controller, and computing an error bound (to solve problem \ref{prob:controller}), the planner states are treated as uncertain parameters. Additionally, from the second point of Theorem~\ref{th:FbL_result}, we know that these uncertain parameters are bounded:  $\mathbf{\hat{x}}(t) \in \hat{X} \ \forall t \in [0, \text{hrz}(\formula)]$, where $\hat{X} = \Re^3 \times [v_\text{min},v_\text{max}]^3 \times [a_\text{min},a_\text{max}]^3$. With this information in mind, the following proposition proposes a way of computing a tracking controller that ensures boundedness of the tracking error. 

\begin{prop}
	Given error dynamics \eqref{eq:err_dyn} with mappings $f_e : \Re^{n}  \times \Re^{\hat{n}}  \rightarrow \Re^{n}$, $g_e : \Re^{n} \rightarrow \Re^{n \times m}$, and $\eta \in \Re$, $\hat{X} \subseteq \Re^{\hat{n}}$, $\underline{u}, \overline{u} \in \Re^{m}$, if there exist a $\mathcal{C}^1$ function $V : \Re^{n} \rightarrow \Re$, and $k : \Re^{n}   \rightarrow \Re^{m}$, such that for all $\hat{x} \in \hat{X} $, the following constraints hold,
	{\small
	\begin{subequations}\label{eq:barrier_cond}
	\begin{align}
	&V(e) = \eta \ \Rightarrow \ \frac{\partial V}{\partial e} \cdot (f_e(e,\hat{x}) + g_e(e)k(e)) \leq 0, \\
	&V(e) \leq \eta \ \Rightarrow \ \underline{u} \leq  k(e) \leq \overline{u},
	\end{align}
	\end{subequations}}
	\hspace{-3pt}then the sublevel set $\Omega_\eta^V:= \{e \in \Re^{n}: V(e) \leq \eta \}$ is forward invariant.
\end{prop}

The sublevel set $\Omega_{\eta}^V$ is the  error bound achieved by the tracking controller $k$. In practice, it can be difficult to find generic functions $V$ and $k$ satisfying constraints \eqref{eq:barrier_cond}. Therefore, we use sum-of-squares (SOS) programming to search for them by restricting them to polynomials: $V \in \Re[e]$, and $k \in \Re^m[e]$. This in turn requires the error dynamics to be polynomials: $f_e \in \Re^n[(e,\hat{a})]$, $g_e \in \Re^{n \times m}[e]$, where $\hat{a} := [\hat{a}_x, \hat{a}_y, \hat{a}_z]^\top$. Additionally, define polynomials $d_l = (a_{\text{max}} - \hat{a}_l)(\hat{a}_l + a_{\text{min}})$, $l \in \{x, y, z\}$.

By applying the generalized S-procedure \cite{Parrilo:00} to \eqref{eq:barrier_cond} to get its corresponding SOS constraints, and using the volume of the error bound as the cost function, we obtain the following SOS optimization problem.

{\small
\begin{subequations} \label{eq:sosopt}
\begin{align}
\min_{V, k, s_i}& \text{volume}(\Omega_{\eta}^V) \nonumber \\
\text{s.t.} \ & s_1^l \in \Sigma[(e,\hat{a})], \ \forall  l \in \{x,y,z\}, \ s_2 \in \Re[(e,\hat{a})],  \nonumber \\
&s_3^j, s_4^j \in \Sigma[e], \ \forall j \in \{1,...,m\},  \nonumber \\
& k \in \Re^m[e], \ V \in \Re[e], \label{eq:A3} \\
&-\frac{\partial V}{\partial e}\cdot (f_e +g_e k)  - \sum_{l \in \{x,y,z\}} s^l_1 d_l \nonumber \\
& \quad - (V-\eta) s_2 \in \Sigma[(e,\hat{a})], \label{eq:A4} \\
&\overline{u}_i - k_i + (V - \eta)s_3^j \in \Sigma[e],\ \forall j \in \{1,...,m\}, \label{eq:A5} \\
&k_i - \underline{u}_i+ (V - \eta)s_4^j   \in \Sigma[e], \ \forall j \in \{1,...,m\}, \label{eq:A6}
\end{align}
\end{subequations}}
\hspace{-7pt}where polynomials decision variables $s_i$ are called S-procedure certificates. The optimization \eqref{eq:sosopt} is non-convex as there are two groups of decision variables $V$ and $(k, s_2, s_3^j, s_4^j)$ bilinear in each other. We tackle it by using the so called alternating direction algorithm summarized in \cite[Algorithm 1]{smithCDC}. As a concrete example of the outputs of optimization~\eqref{eq:sosopt}, the tracking controller for $u_x$ (in \eqref{eq:8state_system}) is:

\vspace{-10pt}
{\small
\begin{align}
u_x(e) =&  
0.001 e_1 e_6 + 0.083 e_2 e_3 
+ 0.139 e_2 e_6 + 0.070 e_3 e_5 \nonumber \\
& - 0.186 e_3 e_7 + 0.003 e_3 e_8 
+ 0.001 e_4 e_6 + 0.120 e_5 e_6 \nonumber \\
&- 0.062 e_6 e_7 + 0.001 e_6 e_8 
- 0.001 e_1 - 0.289 e_2 \nonumber \\
&- 0.002 e_4 - 0.320 e_5 
- 1.142 e_7 + 0.001 e_8. \nonumber
\end{align}}
Once the error bound $\Omega_\eta^V$ is obtained, we can compute the worst-case tracking error $\delta$ by solving a convex optimization
{\small
\begin{align} \label{eq:delta_compute}
&\min_{\delta > 0} \ \text{s.t.} \ \Omega_\eta^V \Rightarrow -\delta \leq e_i \leq \delta, \ i = 1,...,n.
\end{align}}
\vspace{-10pt}

\noindent \textbf{Key result:}  The worst-case tracking error $\delta$ acts as the interface between the control and planning algorithms. Its use is formally stated below:
\begin{theorem} \label{th:main_result}
	Given kinematic constraint set $\hat{X}$,   resulting error bound $\Omega_\eta^V$, tacking controller $k$, and worst-case tracking error $\delta$, if there exists a feasible planned trajectory satisfying the STL specification $\formula$ when solving the optimization \eqref{eq:HLOptim} with $\hat{X}$ and $\delta$, then the trajectories of the nonlinear UAV dynamics \eqref{eq:8state_system} under the control of $k$ satisfy $\formula$.
\end{theorem}

This follows from the controller synthesis procedure, and the trajectory planning guarantees of Theorem \ref{th:FbL_result}. 
The co-design framework ensures that planned trajectories satisfy the STL specification $\formula$, and are dynamically feasible for the controller $k$ to track without violating $\formula$. Hence, together the planner and controller serve as a solution to problem \ref{prob:overall}\footnote{The control synthesis does not take state constraint $X$ into account directly, however it guarantees $\mathbf{x}(t) \in G\hat{X} \oplus \mathbb{B}_\delta$ where $\mathbb{B}_\delta$ is a box with sides $\delta$, $G$ projects $\hat{X}$ on the state space $x$ and $\oplus$ is the Minkowski sum.}.

\noindent \textbf{Co-design procedure:} We now summarize the co-design process, which is outlined in figure~\ref{fig:codesign}. 
\begin{enumerate}
	\item Given the kinematic constraint set $\hat{X}$ and the UAV input constraint set $U$, solve optimizations \eqref{eq:sosopt} and \eqref{eq:delta_compute} to get $k$ and $\delta$ (offline). 
	\item Solve the optimization \eqref{eq:HLOptim} to obtain planned trajectories $\mathbf{\hat{x}}$ satisfying $\formula$ with robustness above $\delta$ (offline).
	\item Track $\mathbf{\hat{x}}$ for each UAV using $k$ (online).
\end{enumerate}

\section{Simulations}
\label{sec:simulations}

We demonstrate our framework through two simulation examples, adapted from \cite{pant2018fly}. The only other approach that can handle the full STL semantics, as well as non-linear dynamics is \cite{pant2017smooth}. However these examples, especially the second example with 4 UAVs, are well beyond its computational abilities. We do not perform comparisons to other approaches as they either are for planning only \cite{pant2018fly, SahaRSJ14}, or cannot handle non-linear dynamics \cite{Raman14_MPCSTL, Sadraddini15Allerton}. 

\noindent \textbf{Simulation setup:} The trajectory planning algorithm and the controller synthesis was implemented in Matlab, and non-linear UAV dynamics \eqref{eq:8state_system} were used for the trajectory tracking. The trajectory planning optimization is implemented using Casadi, and solved using IPOPT as the solver. The trajectory computation times are of the order of a few seconds for both examples. A thorough evaluation of computation times can be found in \cite{pant2018fly}\footnote{The additional constraints for bounded jerk add little to no overhead}. The SOS optimization~\eqref{eq:sosopt} is formulated and translated into semi-definite programming (SDP) using SOSOPT, and the SDP is solved using MOSEK. Variables $V$ and $k$ are parametrized using degree-2 polynomials in both examples, and the corresponding SOS optimization takes $1.75 \times 10^3$ seconds to solve (offline).

\subsection{Two-UAV Timed Reach Avoid}
\label{sec:reachavoid}

We start with the mission in Example \ref{ex:reach_avoid_exmp}, where two UAVs are tasked with a timed reach-avoid mission (see figure \ref{fig:RA}).




\textbf{Results:} In the instance shown in figure \ref{fig:RA}, the UAV initial states correspond to UAV 1 and 2 with positions $[-5,0,2.75]^\top$ and $[-5,2,4]^\top$ respectively, with zero initial velocities and orientations. The trajectory planner generated trajectories of robustness $\rho_{\formula_\text{reach-avoid}}(\mathbf{\hat{x}}) =  1.88$. The synthesized controller has a worst case tracking error (in the inf-norm sense) of $\delta = 1.75$, implying the trajectories can be tracked without violating \eqref{eq:timed_RA}. $\Omega_\eta^V$ is shown with the blue ellipsoids.

%

\begin{figure}[tb]
\centering
\includegraphics[width=0.49\textwidth, trim={1cm 0.5cm 1cm 0cm},clip]{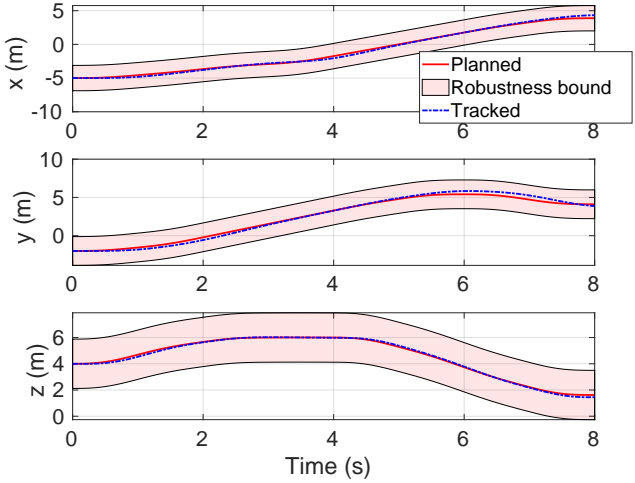}
\vspace{-10pt}
\caption{\small{Position vs time for planned trajectories of UAV 2 flying the timed reach-avoid mission, and after tracking them via the synthesized controller. The actual trajectories (tracked) are within the robustness bound of the planned trajectories, hence ensuring that the specification is satisfied. }}
\label{fig:tracking}
\vspace{-5pt}
\end{figure}

\begin{figure}[tb]
\centering
\includegraphics[width=0.49\textwidth, trim={1cm 0.75cm 1cm .1cm},clip]{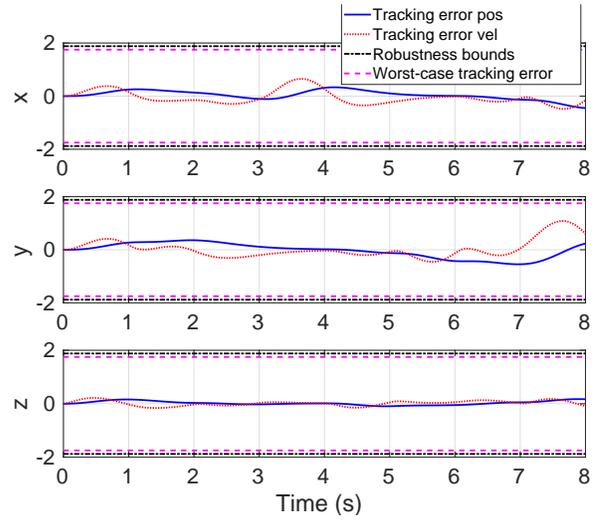}
\vspace{-10pt}
\caption{\small{Tracking errors for position and velocity in the 3 axes of motion for UAV 2 (reach-avoid). Note that the errors are contained within the worst-case bounds, which in turn are smaller than the bounds imposed by the STL robustness of the trajectories.}}
\label{fig:trackingerrors}
\vspace{-5pt}
\end{figure}

Figure \ref{fig:tracking} shows the planned and tracked positions for UAV 2, and figure \ref{fig:trackingerrors} shows the tracking errors, demonstrating that the tracking is indeed good enough to ensure satisfaction of the underlying STL specification.
 This holds for both UAVs, for tracking of both positions and velocities, however we do not present additional figures due to lack of space. A video playback of the simulation can be found at \protect\url{http://bit.ly/TimedRA}

\subsection{Four-UAV Multi-mission example}
\label{sec:case}

Here, the four UAVs are tasked with performing two types of mission in the workspace shown in figure \ref{fig:case}. Two UAVs (1 and 2) have to perform a patrolling mission, twice visiting $\text{Zone 1}$ and $\text{Zone 2}$ between pre-defined time intervals. The other two UAVs (3 and 4) have to visit a $\text{Deliver}$ region within the first two seconds to drop off a package, and then reach the $\text{Base}$ region. For safety, all UAVs must avoid two $\text{Unsafe}$ regions and maintain pairwise separation of at least $0.2$m. This mission has a horizon of $20$ seconds, and is represented in STL as: 

\begin{figure}[tb]
\centering
\includegraphics[width=0.49\textwidth, trim={2cm 1cm 2cm 2cm},clip]{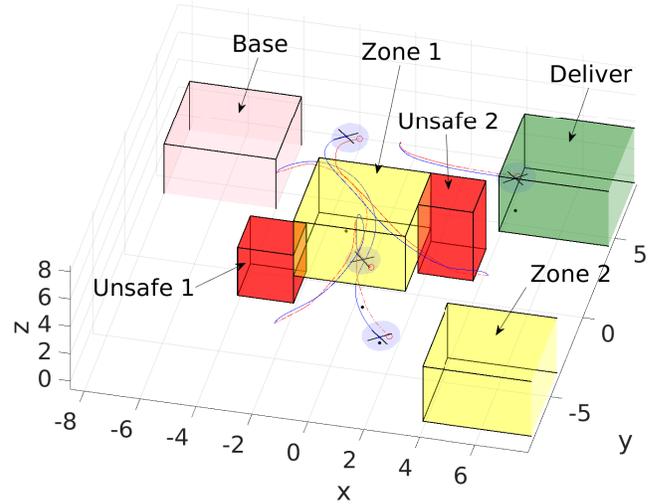}
\vspace{-10pt}
\caption{\small{Workspace for the four UAV multi-mission example. Simulation videos are at \protect\url{http://bit.ly/MultiMission4UAV}}}
\label{fig:case}
\vspace{-20pt}
\end{figure}

{\small
\begin{equation*}
\label{eq:spec_case}
\begin{split}
\formula_\text{multi-mission} &= \land_{i=1}^2 (\eventually_{[0,5]} (p_i \in \text{Zone 1}) \land \eventually_{[5,10]} (p_i \in \text{Zone 2}) \\
&\land \eventually_{[10,15]} (p_i \in \text{Zone 1}) \land \eventually_{[15,20]} (p_i \in \text{Zone 2})) \land \\ & \land \land_{i=3}^4 (\eventually_{[0,10]} (p_i \in \text{Deliver}) \land \eventually_{[10,20]} (p_i \in \text{Base})) \land \\ &\land \land_{i=1}^4 (\always_{[0,20]} (\neg p_i \in \text{Unsafe}_1) \land \always_{[0,20]} (\neg p_i \in \text{Unsafe}_2))  \\ & \land_{i,j,i \neq j} \always_{[0,20]} (||p_i-p_j|| \geq 0.2)
\end{split}
\end{equation*}}
\vspace{-5pt}

\textbf{Results:} The trajectory planner generates trajectories that satisfy the specifications with a robustness $\rho_{\formula_\text{multi-mission}}(\mathbf{\hat{x}})=1.85 \geq \delta = 1.75$, ensuring that the trajectories can be tracked without violating the specification. $\Omega_\eta^V$ is shown with the blue ellipsoids in figure \ref{fig:case}. The planned and tracked trajectories are shown with red and blue curves, respectively. A video playback of the simulation can be found at \protect\url{http://bit.ly/MultiMission4UAV}
\section{Conclusion}
\label{sec:conclusion}
\noindent\textbf{Summary:} In this paper, we present an approach for planning and control of (multiple) UAVs carrying out tasks specified using Signal Temporal Logic (STL). This is done by co-designing the trajectory generator and the controller that tracks these trajectories. We showed how this co-design allows us to take into account kinematic constraints, the full non-linear UAV dynamics, and the tracking error while generating and flying out trajectories that satisfy the STL specification in continuous time. Through simulations, we show the the co-design results in trajectories that satisfy the STL specification and a synthesized controller that tracks them within prescribed bounds.

\noindent\textbf{Limitations and future work:}
In practice, the computed worst-case tracking bounds for the synthesized controller can be very conservative. This is noticed in the simulations where the tracking error never approaches these bounds. This could result in the trajectory planning optimization being unable to obtain a high enough robustness value with respect to the given STL specification for Theorem \ref{th:FbL_result} to hold, e.g. in cases where the UAVs need to fly through narrow openings, and restricts the application of our approach to such settings. To overcome this, we aim to improve our technique by \textit{tightening} the STL specification by a state-dependent function of the tracking error in the planning phase, instead of using the worst-case bounds on tracking error. Additionally, our hierarchical planning and control framework is limited to specifications only over the translational states of the system and cannot handle temporal specifications on the orientations of the UAVs.


\bibliographystyle{IEEEtran}
\bibliography{root}

\begin{thebibliography}{10}

\bibitem{pant2018fly}
Y.~V. Pant, H.~Abbas, R.~A. Quaye, and R.~Mangharam, ``Fly-by-logic: control of
  multi-drone fleets with temporal logic objectives,'' in {\em Proceedings of
  the 9th ACM/IEEE International Conference on Cyber-Physical Systems},
  pp.~186--197, IEEE Press, 2018.

\bibitem{pant2017smooth}
Y.~V. Pant, H.~Abbas, and R.~Mangharam, ``Smooth operator: Control using the
  smooth robustness of temporal logic,'' in {\em Control Technology and
  Applications, 2017 IEEE Conf. on}, pp.~1235--1240, IEEE, 2017.

\bibitem{8404080}
L.~{Lindemann} and D.~V. {Dimarogonas}, ``Control barrier functions for signal
  temporal logic tasks,'' {\em IEEE Control Systems Letters}, 2019.

\bibitem{SahaRSJ14}
I.~Saha, R.~Rattanachai, V.~Kumar, G.~J. Pappas, and S.~A. Seshia, ``Automated
  composition of motion primitives for multi-robot systems from safe ltl
  specifications,'' in {\em IEEE/RSJ International Conference on Intelligent
  Robots and Systems}, 2014.

\bibitem{Parrilo:00}
P.~Parrilo, ``Structured semidefinite programs and semialgebraic geometry
  methods in robustness and optimization,'' 2000.

\bibitem{Raman14_MPCSTL}
V.~Raman, A.~Donze, M.~Maasoumy, R.~M. Murray, A.~Sangiovanni-Vincentelli, and
  S.~A. Seshia, ``Model predictive control with signal temporal logic
  specifications,'' in {\em 53rd IEEE Conf. on Decision and Control},
  pp.~81--87, Dec 2014.

\bibitem{Sadraddini15Allerton}
S.~Sadraddini and C.~Belta, ``Robust temporal logic model predictive control,''
  in {\em Allerton conference}, September 2015.

\bibitem{bemporad1999control}
A.~Bemporad and M.~Morari, ``Control of systems integrating logic, dynamics,
  and constraints,'' {\em Automatica}, vol.~35, no.~3, pp.~407--427, 1999.

\bibitem{kousik2018bridging}
S.~Kousik, S.~Vaskov, F.~Bu, M.~Johnson-Roberson, and R.~Vasudevan, ``Bridging
  the gap between safety and real-time performance in receding-horizon
  trajectory design for mobile robots,'' {\em arXiv preprint arXiv:1809.06746},
  2018.

\bibitem{herbert2017fastrack}
S.~L. {Herbert}, M.~{Chen}, S.~{Han}, S.~{Bansal}, J.~F. {Fisac}, and C.~J.
  {Tomlin}, ``Fa{ST}rack: A modular framework for fast and guaranteed safe
  motion planning,'' in {\em 2017 IEEE 56th Annual Conference on Decision and
  Control (CDC)}, pp.~1517--1522, Dec. 2017.

\bibitem{Singh2018RobustTW}
S.~Singh, M.~Chen, S.~L. Herbert, C.~J. Tomlin, and M.~Pavone, ``Robust
  tracking with model mismatch for fast and safe planning: an sos optimization
  approach,'' {\em arXiv preprint arXiv:1808.00649}, 2018.

\bibitem{GIRARD_CA}
A.~Girard and G.~J. Pappas, ``Hierarchical control system design using
  approximate simulation,'' {\em Automatica}, vol.~45, no.~2, pp.~566 -- 571,
  2009.

\bibitem{SMITH_automatica}
S.~W. Smith, M.~Arcak, and M.~Zamani, ``Approximate abstractions of control
  systems with an application to aggregation,'' {\em Automatica}, vol.~119,
  p.~109065, 2020.

\bibitem{smithCDC}
S.~W. {Smith}, H.~{Yin}, and M.~{Arcak}, ``Continuous abstraction of nonlinear
  systems using sum-of-squares programming,'' in {\em 2019 IEEE 58th Conference
  on Decision and Control (CDC)}, pp.~8093--8098, 2019.

\bibitem{PJ_IFAC}
P.-J. {Meyer}, H.~{Yin}, A.~H. {Brodtkorb}, M.~{Arcak}, and A.~J.
  {S{\o}rensen}, ``{Continuous and discrete abstractions for planning, applied
  to ship docking},'' {\em arXiv e-prints}, p.~arXiv:1911.09773, Nov. 2019.

\bibitem{YinAcc}
H.~{Yin}, M.~{Bujarbaruah}, M.~{Arcak}, and A.~{Packard}, ``Optimization based
  planner–tracker design for safety guarantees,'' in {\em 2020 American
  Control Conference (ACC)}, pp.~5194--5200, 2020.

\bibitem{MalerN2004STL}
O.~Maler and D.~Nickovic, {\em Monitoring Temporal Properties of Continuous
  Signals}.
\newblock Springer Berlin Heidelberg, 2004.

\bibitem{FainekosP09tcs}
G.~Fainekos and G.~Pappas, ``Robustness of temporal logic specifications for
  continuous-time signals,'' {\em Theor. Computer Science}, 2009.

\bibitem{sabatino2015quadrotor}
F.~Sabatino, ``Quadrotor control: modeling, nonlinearcontrol design, and
  simulation,'' 2015.

\bibitem{MuellerTRO15}
M.~W. Mueller, M.~Hehn, and R.~D\'Andrea, ``A computationally efficient motion
  primitive for quadrocopter trajectory generation,'' in {\em IEEE Transactions
  on Robotics}, 2015.

\end{thebibliography}
\appendix
\label{sec:appendix}

For the constraints of the trajectory planning optimization \eqref{eq:HLOptim}, consider the min-jerk trajectory segment \cite{MuellerTRO15}, of time duration $T$, between a waypoint  $\hat{p}^j=[\hat{p}^j_x,\hat{p}^j_y,\hat{p}^j_z]^\top$ with velocity $\hat{v}^j = [ \hat{v}^j_x, \hat{v}^j_y, \hat{v}^j_z]^\top$, and the next waypoint with desired position $\hat{p}^{j+1} =[\hat{p}^1_x,\hat{p}^1_y,\hat{p}^1_z]^\top$. We define the functions \cite{pant2018fly}, for $t \in [0,T]$:


\vspace{-10pt}
{\small
\begin{equation}
\begin{split}
K_3(t) &= (90t^4)/(48T^5) - (90t^3)/(12T^4)+(30t^2)/(4T^3) \\
 K_4(t) &= (90t^3)/(12T_f^5) - (90t^2)/(4T^4)+(30t)/(2T^3)
\end{split} 
\end{equation}}
\vspace{-10pt}

Let $t' = \argmax_{t \in [0,T]} K_t(t)$. We can now define the constraints (for each UAV) that ensure velocity and acceleration are within bounds $[v_\text{min},v_\text{max}]$ and $[a_\text{min},a_\text{max}]$ respectively, for each axis of motion $l$:


\vspace{-10pt}
{\small
\begin{equation}
\begin{split}
\text{LB}_\text{vel}(\hat{v}^j_l) &= (v_\text{min}-(1-TK_3(T)) \hat{v}^j_l)/ K_3(T)  \\
\text{UB}_\text{vel}(\hat{v}^j_l) &= (v_\text{max}-(1-TK_3(T)) \hat{v}^j_l )/K_3(T) \\
\text{LB}_\text{acc}(\hat{v}^j_l) &= T\hat{v}^j_l + a_\text{min}/K_4(t') \\
\text{UB}_\text{acc}(\hat{v}^j_l) &= T\hat{v}^j_l + a_\text{max}/K_4(t')
\end{split} 
\end{equation}}
\vspace{-10pt}

Combining these constraints for all axis of motion gives $x, y,z$ the velocity and acceleration constraints in the optimization \eqref{eq:HLOptim} of the form $\text{LB}_\text{vel}(\hat{v}^j) = [\text{LB}_\text{vel}(\hat{v}^j_x), \text{LB}_\text{vel}(\hat{v}^j_y),\text{LB}_\text{vel}(\hat{v}^j_z)]$ and similarly for the upper bound for velocities and upper/lower bounds for acceleration. These constraints are such that:

\vspace{-10pt}
{\small
\begin{equation}
\begin{split}
& \text{LB}_\text{vel}(\hat{v}^j) \leq \hat{p}^{j+1} - \hat{p}^j \leq  \text{UB}_\text{vel}(\hat{v}^j) \\ &\Rightarrow \hat{v}^j_l \in [v_\text{min},v_\text{max}]\, \forall t \in [0,T], \, \forall l \in \{x,y,z\}, \,\text{and,} \\
& \text{LB}_\text{acc}(\hat{v}^j) \leq \hat{p}^{j+1} - \hat{p}^j \leq  \text{UB}_\text{acc}(\hat{v}^j) \\ &\Rightarrow \hat{a}^j_l \in [a_\text{min},a_\text{max}]\, \forall t \in [0,T], \, \forall l \in \{x,y,z\} 
\end{split}
\end{equation}
}

Finally, the constraint that ensures bounded jerk $\mathbf{\hat{j}}_l(t) \in [j_\text{min},j_\text{max}]\, \forall t \in[0,T]$ for each axis of motion $l$ are:

\vspace{-10pt}
{\small
\begin{equation*}
\begin{split}
\text{LB}_\text{jerk}(\hat{v}^j_l) &= \min((2T^3/30)j_\text{min}-0.5T\hat{v}^j_l,\, (2T^3/30)j_\text{min}+T\hat{v}^j_l ) \\
\text{UB}_\text{jerk}(\hat{v}^j_l) &= \min((2T^3/30)j_\text{max}-0.5T\hat{v}^j_l,\, (2T^3/30)j_\text{max}+T\hat{v}^j_l ) 
\end{split} 
\end{equation*}
}
\vspace{-10pt}

The proof follows a similar construction as the proofs in \cite{pant2018fly}. Note, these constraints are linear in the position waypoints, which are the variables of the trajectory planning optimization \eqref{eq:HLOptim}.

\end{document}